\documentclass[hyper,letterpaper,12pt]{article}
\usepackage{a4wide}
\usepackage{amsmath}
\usepackage[
      colorlinks=true,
      linkcolor=blue,
      urlcolor=blue,
      filecolor=black,
      citecolor=blue,
            ]{hyperref}
\usepackage{color}
\usepackage{graphicx}
\usepackage{amsfonts}
\usepackage{amssymb}
\usepackage{epsfig}
\usepackage{subfigure}
\usepackage{amsmath}
\usepackage{latexsym}
\usepackage{mathrsfs}

\begin{document}
\title{\bf \Large Holographic Van der Waals-like phase transition in the Gauss-Bonnet gravity}

\author{\large
~Song He$^{3,1}$\footnote{E-mail: hesong17@gmail.com}~,
~~Li-Fang Li$^4$\footnote{E-mail: lilf@itp.ac.cn}~,
~~Xiao-Xiong Zeng$^{1,2}$\footnote{E-mail: xxzeng@itp.ac.cn}~
\date{\today}
\\
\\
\small $^1$State Key Laboratory of Theoretical Physics, Institute of Theoretical Physics,\\
\small Chinese Academy of Sciences, Beijing 100190,  China\\
\small $^2$ School of Material Science and Engineering, Chongqing Jiaotong University,\\
\small       Chongqing ~400074, China\\
\small $^3$ Max Planck Institute for Gravitational Physics (Albert Einstein Institute),\\
\small Am M\"{u}hlenberg 1, 14476 Golm, Germany\\
\small $^4$ Center for Space Science and Applied Research, Chinese Academy of Sciences,\\
\small Beijing 100190, China\\}

\maketitle

\begin{abstract}
\normalsize The Van der Waals-like phase transition is observed in temperature-thermal entropy plane in spherically symmetric charged Gauss-Bonnet-AdS black hole background. In terms of AdS/CFT, the non-local observables such as holographic entanglement entropy, Wilson loop, and two point correlation function of very heavy operators in the field theory dual to spherically symmetric charged Gauss-Bonnet-AdS black hole have been investigated.  All of them exhibit the Van der Waals-like phase transition for a fixed charge parameter or Gauss-Bonnet parameter in such gravity background. Further, with choosing various values of charge or Gauss-Bonnet parameter, the equal area law and the critical exponent of the heat capacity are found to be consistent with phase structures in temperature-thermal entropy plane.
\end{abstract}

\newpage

\tableofcontents

\section{Introduction}
The Van der Waals-like  behavior of a black hole is an interesting phenomenon in black hole physics. It helps us to understand new phase structure in black hole thermodynamics. In the pioneering work~\cite{Chamblin}, it was found that a charged AdS black hole exhibits the Van der Waals-like phase transition in the $T-S $ plane. As the charge of the black hole increases from small to large, the black hole will undergo first order phase transition and second order phase transition successively before it reaches to a stable phase, which is analogous to the van der Waals liquid-gas phase transition. The Van der Waals-like phase transition has also been observed in the $Q-\Phi$ plane \cite{Niu}, where $Q$ is electric charge and $\Phi$ is the chemical potential. Further, the Van der Waals-like phase transition can be realized in the $P-V$ plane \cite{Kubiznak,Xu,Caipv,Hendi,Hennigar,Wei1,Mo,Spallucci}. Where the negative cosmological constant is treated as the  pressure $P$ and the thermodynamical volume $V$ is the conjugating quantity of pressure.

By AdS/CFT \cite{ads1,ads2,ads3}, \cite{Johnson} has investigated holographic entanglement entropy \cite{Ryu:2006bv,Ryu:2006ef} in a finite volume quantum system which is dual to a  spherical and charged $AdS_4$ black hole. Their results showed that there exists Van der Waals-like phase transition in the entanglement entropy-temperature plane. This phase transition is analogy with thermal dynamical phase transition. The critical exponent of the heat capacity for the second order phase transition was found to be consistent with that in the mean field theory. Meanwhile \cite{Nguyen} investigated exclusively the equal area law in the entanglement entropy-temperature plane and found that the equal area law  holds  regardless of the size of the entangling region. There have been some extensive studies \cite{Caceres,zeng20161,zeng20162,zeng20163,Dey,mo1} and all the results showed that as the case of thermal dynamical entropy, the entanglement entropy  exhibited the Van der Waals-like phase transition. These results indicate that there are some intrinsic relation between black hole entropy and holographic entanglement entropy. Furthermore, expectation value of Wilson loop \cite{Li:2011hp,Cai:2012xh,Cai:2012sk,Cai:2012nm,Cai:2012es} and the equal time two point correlation function of heavy operators have some similar properties as the entanglement entropy \cite{Balasubramanian1,Balasubramanian2,GS,CK, Zeng20151,Liuh,Zhangs,Buchel, Craps,Camilo} to reveal the phase transitions in quantum systems.

In this paper, we would like to extend ideas in \cite{Johnson} to study van der Waals-like phase transitions in a Gauss-Bonnet-AdS black hole with a spherical horizon in (4+1)-dimensions in the framework of holography. Firstly, we observe that the thermal dynamical entropy will undergo the Van der Waals-like phase transition in temperature-thermal entropy plane. We also study Maxwell's equal area law and critical exponent of the heat capacity, which are two characteristic quantities in van der Waals-like phase transition. Secondly, we would like to study the  holographic entanglement entropy for  a fixed size of entangled region to confirm whether  there is Van der Waals-like phase transition. More precisely speaking, considering that the  holographic entanglement entropy formula should have quantum correction when the bulk theory have higher curvature
terms. In terms of \cite{deBoer:2011wk,Hung:2011xb,Fursaev:2006ih, Ogawa:2011fw,Myers:2010xs,Dong:2013qoa,Camps:2013zua,Miao:2014nxa},
one can study the  holographic entanglement entropy with higher derivative gravity and see what
will happen for the entanglement entropy. Further, we study the expectation value of Wilson loop and two point correlation function of heavy operator in the dual field theory to check whether these two objects also undergo the Van der Waals-like phase transition. We also check  the analogous equal area law and critical exponent of the analogous heat capacity, which are to make sure that all these nonlocal quantum observables will undergo van der Waals-like phase transition in the field theory dual to spherical Gauss-Bonnet-AdS black holes. Our results  confirm the fact that the nonlocal quantum objects are good quantities to probe the phase structures of the spherical Gauss-Bonnet-AdS black holes.

Our paper is organized as follows. In section 2, we review the black hole
thermodynamics for the spherically symmetric Gauss-Bonnet-AdS black hole and discuss the Van der Waals-like phase transition in the $T-S $ plane.
We also check Maxwell's equal area law and critical exponent of the heat capacity numerically. In section 3, with the
 holographic entanglement entropy, Wilson loop, and two point correlation function, we will show all these quantum objects undergo Van der Waals-like phase transition in the spherical Gauss-Bonnet-AdS black hole. In each subsection, the equal area law is checked and the critical exponent of the analogues heat capacity is obtained via data fitting. In the final section, we present our conclusions.

\textbf{Note Added}: While this paper was close to completion, we find \cite{Sun}
also investigate holographic phase transition for a neutral Gauss-Bonnet-AdS black hole in the extended phase space, which
partially overlaps with our work.

\section{Thermodynamic phase transition in the Gauss-Bonnet gravity}
\subsection{Review of the Gauss-Bonnet-AdS black hole}
\label{quintessence_Vaidya_AdS}
The 5-dimensional Lovelock gravity can be realized by adding the
Gauss-Bonnet term to pure Einstein gravity theory.
As a matter field is considered, the theory can be described by the following action \cite{0912.1944}
\begin{eqnarray}
I = \frac{1}{2{\ell_p}^3} \int \mathrm{d}^5x \, \sqrt{-g}\, \left[ R+
\frac{12}{L^2}  + \frac{\alpha L^2}{2} L_4-4 \pi F_{\mu\nu}^{\mu\nu}  \right],
 \end{eqnarray}
with
\begin{eqnarray}
 L_4=R_{\mu\nu\rho\sigma}R^{\mu\nu\rho\sigma}-4R_{\mu\nu}R^{\mu\nu}+R^2,
\end{eqnarray} where ${\ell_p}$ is Newton constant, $\alpha$ denotes the coupling
of Gauss-Bonnet gravity, $L$ stands for the Radius of AdS
background, which satisfies the relation $L^2 = - \frac{6}{\Lambda}$, $F_{\mu\nu}=\partial_{\mu}A_{\nu}- \partial_{\nu}A_{\mu} $ is the Maxwell field strength with the vector potential $A_{\mu}$. In this paper, we use geometric units of $c=G=\hbar =k_B=1$. The Gauss-Bonnet-AdS black hole can be written as \cite{Cai,Johnson2,Cvetic}
\begin{eqnarray}\label{peturbation}
ds^{2}=-f(r)dt^2+\frac{dr^2}{f(r)}+r^{2}[d\theta^2+\sin^2\theta (d\phi^2+\sin^2\theta d\psi^2)],
\label{AADS}
\end{eqnarray}
in which
\begin{eqnarray}\label{peturbation}
f(r)=\frac{r^2 }{2 \alpha }\left(1-\sqrt{-\frac{4 \alpha }{L^2}+\frac{32 \alpha  M}{3 \pi  r^4}-\frac{16 \alpha  Q^2}{3 \pi ^2 r^6}+1}\right)+1,
\label{AADS}
\end{eqnarray}
where $M$ is the mass and  $Q$ is the charge of the black hole.  In the low energy effective action of heterotic string theory, $\alpha$
is proportional to the inverse string tension with positive parameter. Thus in this paper
we will consider the case $\alpha>0$ \cite{Buchel, 0911.3160}. In addition,  from  (\ref{peturbation}), one can see  that there is an upper bound for the
Gauss-Bonnet parameter, namely $\alpha < L^2/4$ .

In the Gauss-Bonnet-AdS background, the black hole event horizon $r_h$ is the largest root of the equation  $f (r_h)=0$.
At the event horizon, the Hawking temperature can be expressed as \cite{Cai2}
\begin{equation}
T=\frac{L^2 \left(3 \pi ^2 r_h^6 \left(\sqrt{\frac{\left(2 \alpha +r_h^2\right)^2}{r_h^4}}-1\right)-8 \alpha  Q^2\right)+12 \pi ^2 \alpha  r_h^6}{12 \pi ^3 \alpha  L^2 r_h^5 \sqrt{\frac{\left(2 \alpha +r_h^2\right)^2}{r_h^4}}},\label{temperature}
 \end{equation}
in which we have used the relation
\begin{equation}
M=\frac{4 L^2 Q^2+3 \pi ^2 L^2 r_h^4+3 \pi ^2 \alpha  L^2 r_h^2+3 \pi ^2 r_h^6}{8 \pi  L^2 r_h^2}.\label{m}
 \end{equation}
The chemical potential in this background is
\begin{equation}
\Phi=\frac{Q}{\pi r_h^2}.\label{cp}
 \end{equation}
The entropy of  the black hole can be written as
\begin{equation}
S=\frac{\pi ^2 r_h^3}{2}  \left(\frac{6 \alpha }{r_h^2}+1\right).\label{cp}
 \end{equation}
Inserting  (\ref{cp}) into  (\ref{temperature}), we can get the relation between the temperature and entropy $T(S,Q)$. Next, we will employ this relation to study the phase structure of the Gauss-Bonnet-AdS   black hole in the $T-S$ plane.

\subsection{Phase transition of thermal entropy}

As we know, for a charged AdS  black hole, the spacetime undergos the Van der Waals-like  phase transition as the charge changes from a small value to a  large value. Especially there is a critical charge, for which the temperature and entropy satisfy the following relation
\begin{equation}
\left(\frac{\partial T}{\partial S}\right)_Q=\left(\frac{\partial^2 T}{\partial S^2}\right)_Q=0. \label{heat1}
 \end{equation}
In our background, the function $T(S,Q)$ is too prolix so that we are hard to get the analytical value of the critical charge. We will get the critical charge   numerically.
In the  Gauss-Bonnet gravity, it has been found that not only the charge but also the  Gauss-Bonnet parameter will affect the phase structure of the black hole. When we discuss the effect of $\alpha$ on the phase structure, the symbol $Q$ in  (\ref{heat1}) should be replaced by $\alpha$.

In order to obtain an analogy with the liquid-gas phase transition in fluids, we can identify free energy $F$ of black hole with the Gibbs free energy $G=G(P,V)$ of the fluid. Where the $P, V$ correspond to pressure and volume of fluid. In \cite{Kubiznak}, they identify cosmology constant and curvature in black hole as pressure and volume to study analogy thermal dynamics.
In \cite{Kubiznak}, $(T,S), (\Phi, Q)   \text{and}    (V, P)$   are interpreted as conjugated variables in AdS black system. For our case, one can turn off $\alpha$ to obtain the AdS black hole system studied in \cite{Kubiznak}. In order to avoid the confusion, we choose two kinds of identifications shown in [\ref{table1}].
\begin{equation}\label{table1}
\begin{array}{|c|c|c|}
\hline
\multicolumn{3}{|c|}{\mbox{Analogy }} \\
\hline
\mbox{fluid} & \mbox{Gauss-Bonnet AdS black hole} & \mbox{Gauss-Bonnet AdS black hole}\\
\hline	
\mbox{temperature} & Q& \alpha \\
\mbox{pressure, P} & T( \alpha )& T( Q)\\
\mbox{volume, V} & S(\alpha)&S( Q)\\
\hline
\end{array}
\end{equation}
\emph{It should be stressed that though the Van der Waals-like phase transition can be constructed by transposing intensive with extensive variables with the help of  (10), the fluid analogy of the Gauss-Bonnet-AdS black hole in our paper is incomplete. We should emphasize that the complete understanding of the fluid analogy and exact definition of extensive variables are given by \cite{Kubiznak}. More precisely, in \cite{Kubiznak}, the cosmological constant is treated as a thermodynamic pressure and its conjugate quantity as a thermodynamic volume. The complete fluid analogy has been presented in \cite{Kubiznak}. In later part of this paper, we are mainly interested in the relation between the thermodynamic entropy and entanglement entropy. Thus it is convenient to discuss the Van der Waals-like phase transition in the $T-S$ plane and we can compare the phase structure of thermodynamic entropy and entanglement entropy transparently.}

\begin{figure}[h!]
\centering
\subfigure[$Q=0$]{
\includegraphics[scale=0.75]{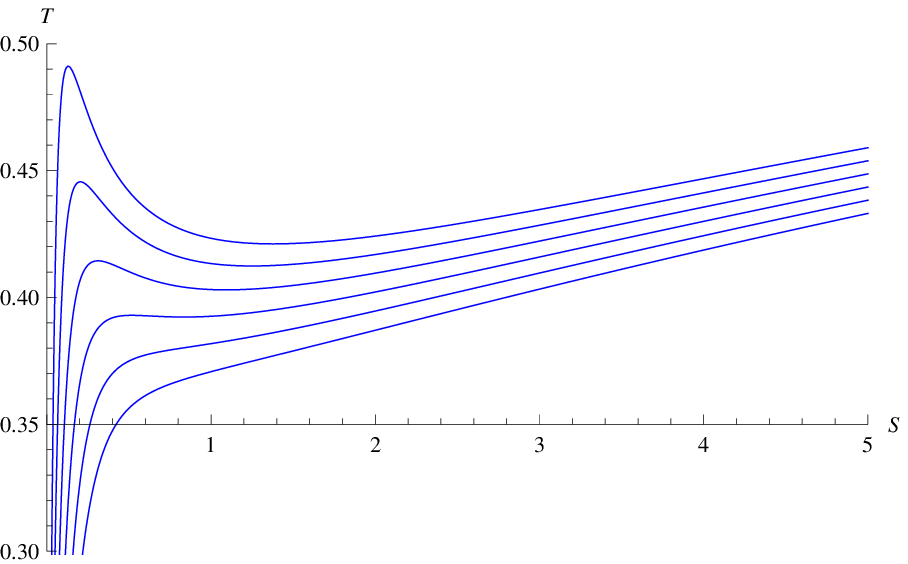}
}
\subfigure[$Q=0$]{
\includegraphics[scale=0.75]{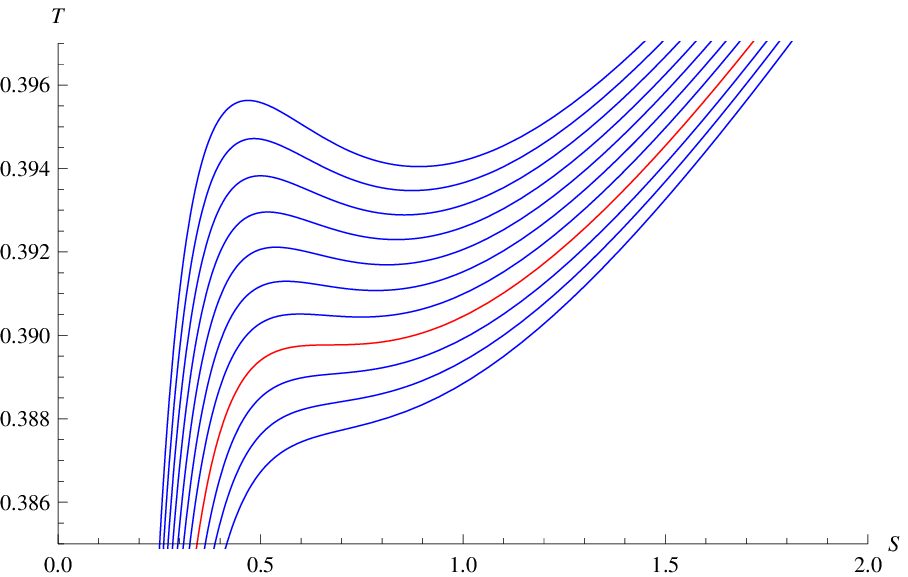}
}
\caption{\label{fig1} \small Relations between entropy and temperature  for different $\alpha$  with a fixed $Q$. In (a), curves from top to down correspond to  cases $\alpha$ varies from 0.015 to 0.035 with step 0.004, while in (b) these curves correspond to  cases $\alpha$ varies from 0.0264 to 0.0284 with step 0.0002.}
\end{figure}

Firstly, we will fix the charge to discuss how the  Gauss-Bonnet parameter  affects the phase structure. We will set $L=1$.
For the case $Q=0$, we know that in the Einstein gravity, the black hole undergos the Hawking-Page transition. But in our background, we find the black hole undergos the Van der Waals-like  phase transition, which is shown in (a) of  Figure  \ref{fig1}.

\begin{figure}[h!]
\centering
\subfigure[$Q=0$]{
\includegraphics[scale=0.6]{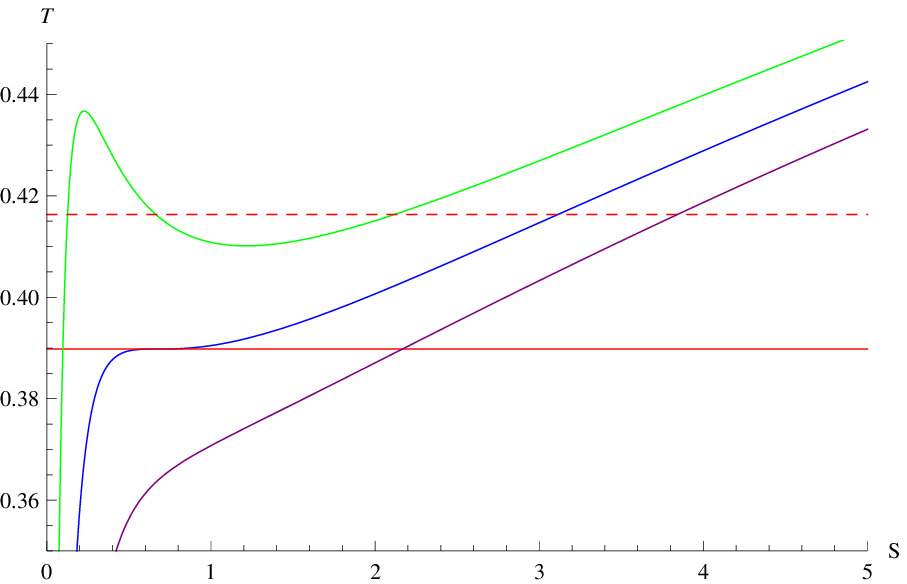}
}
\subfigure[$Q=0.1$]{
\includegraphics[scale=0.6]{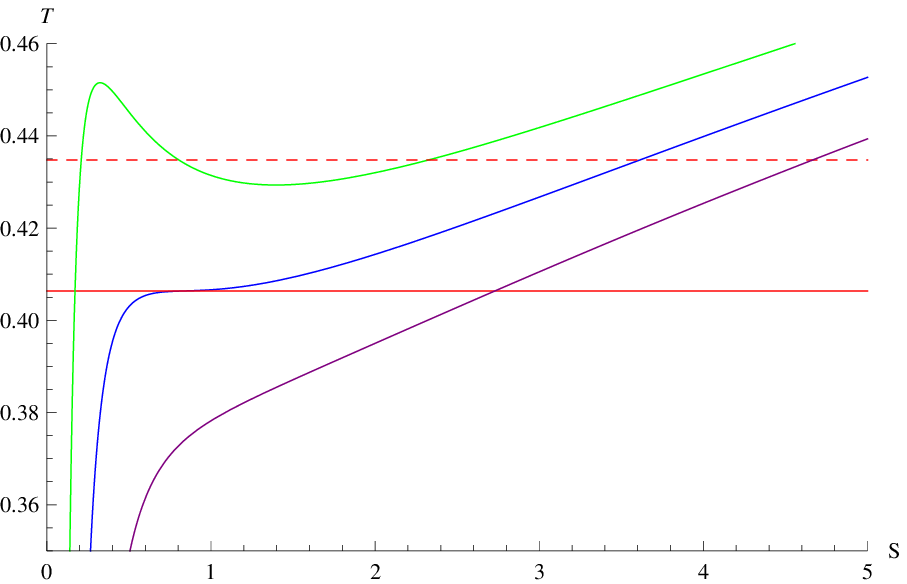}
}
\caption{\label{fig2}\small Relations between entropy and temperature  for different $\alpha$   with a fixed $Q$. In (a), curves from top to down correspond to $\alpha=0.02,0.0277925, 0.035$, and  in (b) they correspond to $\alpha=0.01,0.01972, 0.03$ respectively.
The red   dashed  line and solid line correspond to the first order  phase transition temperature $T_f$ and second order phase transition temperature $T_c$.
 }
\end{figure}

\begin{figure}[h!]
\centering
\subfigure[$Q=0, \alpha=0.02$]{
\includegraphics[scale=0.6]{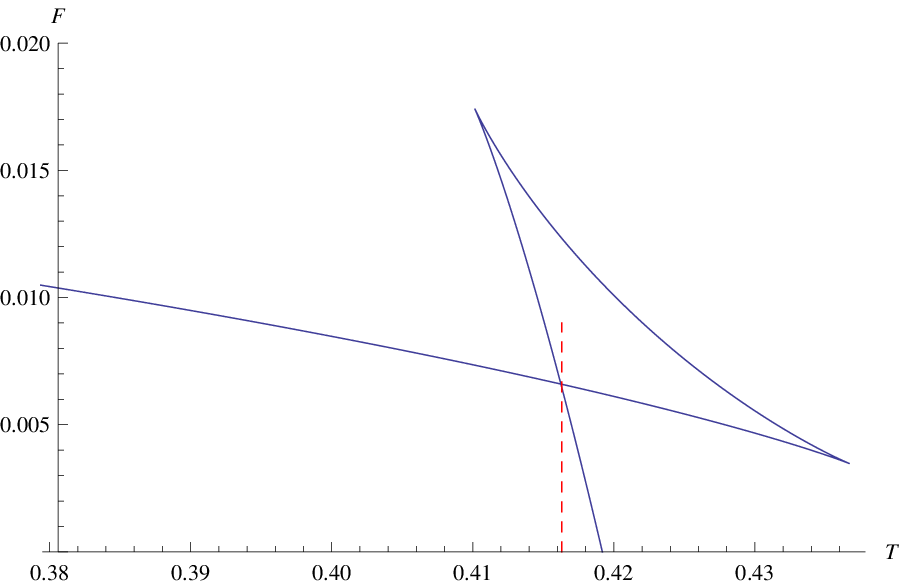}
}
\subfigure[$Q=0.1, \alpha=0.01$]{
\includegraphics[scale=0.6]{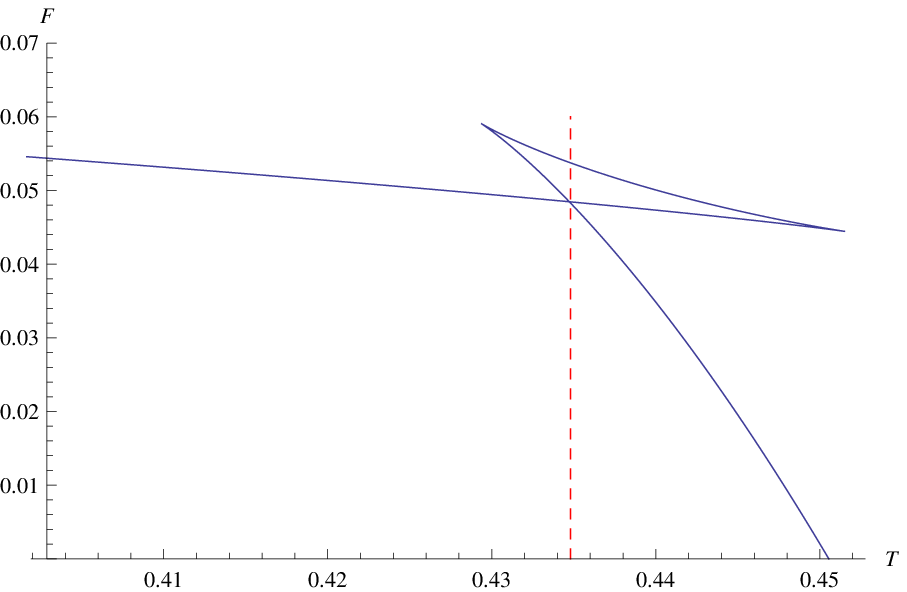}
}
\caption{\label{fig3}\small Relations between the free energy and temperature. }
\end{figure}
\begin{figure}
\centering
\subfigure[$\alpha=0.01$]{
\includegraphics[scale=0.6]{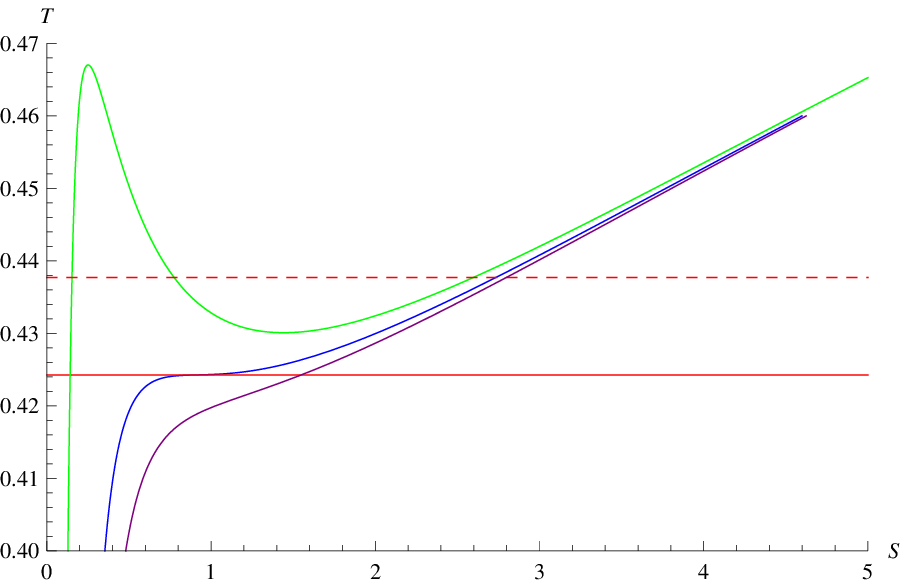}
}
\subfigure[$\alpha=0.02$]{
\includegraphics[scale=0.6]{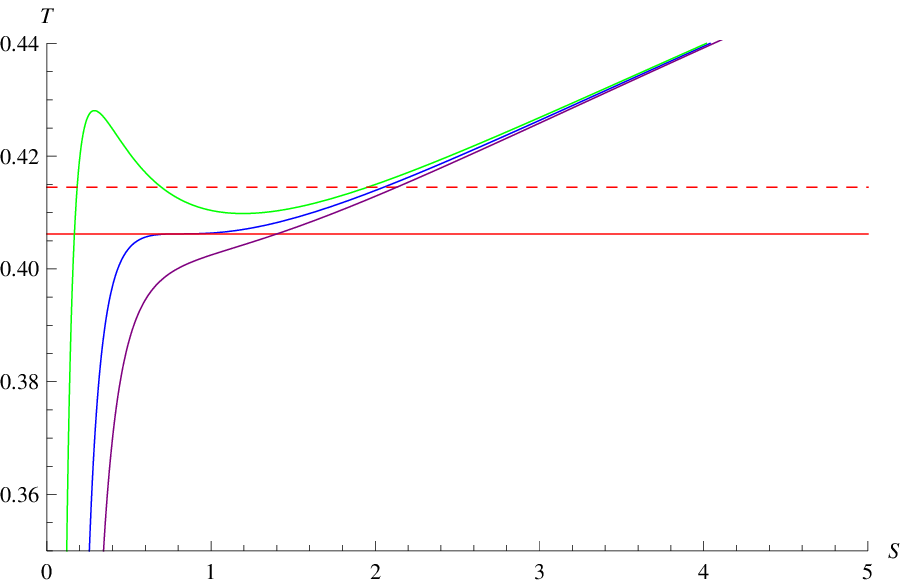}
}
\caption{\small Relations between entropy and temperature  for different $Q$ with a fixed $\alpha$. In (a), curves from top to down correspond to $Q=0.08,0.1681103, 0.2$, and  in (b) they correspond to $Q=0.03,0.094984, 0.13$ respectively. The red dashed  line and solid line correspond to the first order  phase transition temperature $T_f$ and second order phase transition temperature $T_c$.} \label{fig4}
\end{figure}
\begin{figure}[h!]
\centering
\subfigure[$Q=0.08,\alpha=0.01$]{
\includegraphics[scale=0.6]{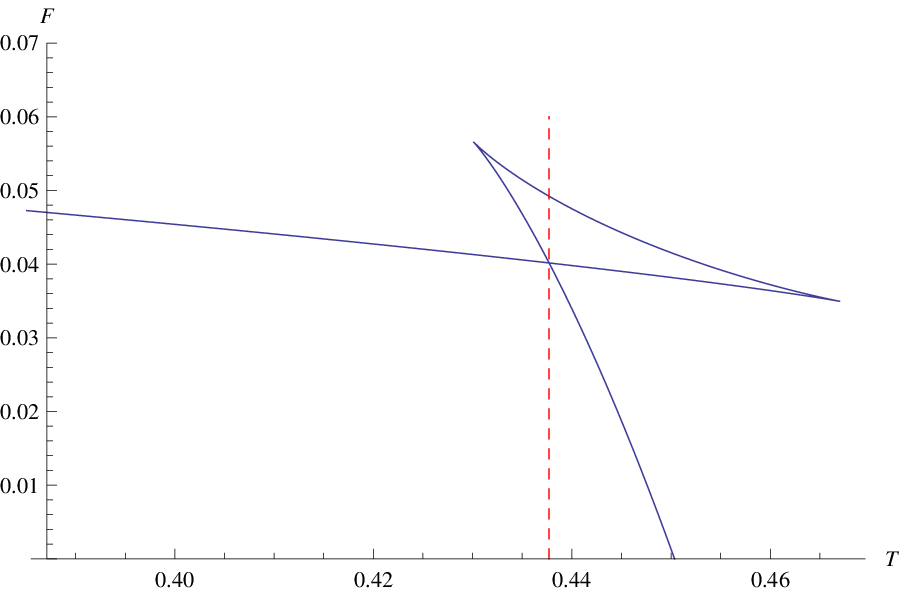}
}
\subfigure[$Q=0.03,\alpha=0.02$]{
\includegraphics[scale=0.6]{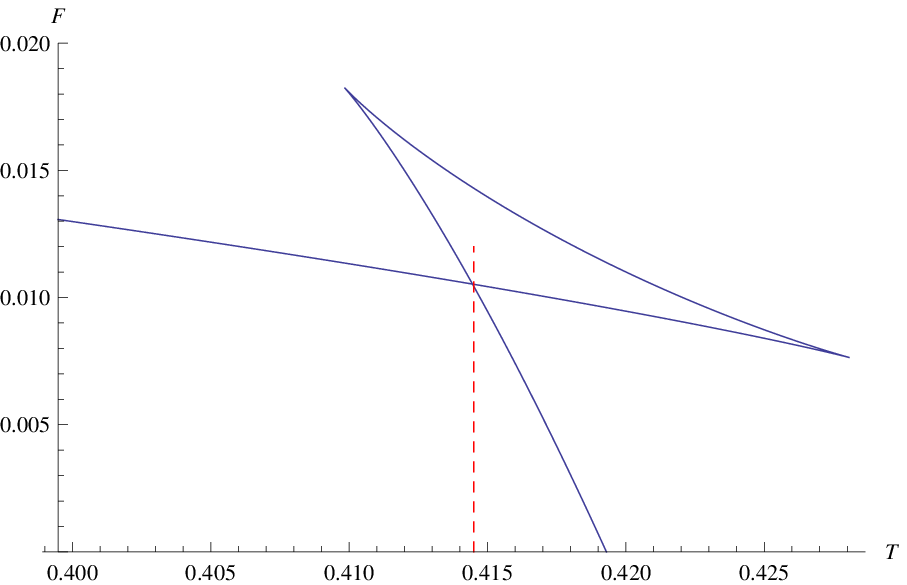}
}
\caption{\label{fig5}\small  Relations between the free energy and temperature. }
\end{figure}
 Exactly, to get the Van der Waals-like  phase transition in this case, we should find  a critical value of the Gauss-Bonnet parameter.  For the function  $T(S,Q)$ is too prolix, we will get it numerically. { We plot a series of curves with taking different values of $\alpha$ in the $T-S$ plane shown in (a) of Figure \ref{fig1}, and one can read off the region of critical value of the Gauss-Bonnet parameter $\alpha$ which satisfies with the condition $(\frac{\partial T}{\partial S})_{\alpha}=0$. We plot a bunch of curves in the $T-S$ plane with smaller step so that we can get the precise critical value of $\alpha$. From (b) of Figure  \ref{fig1}, we find the exact critical value of the Gauss-Bonnet parameter should be about $0.0278$, which are labeled by the red dashed lines in (b) of Figure  \ref{fig1}. Finally, we adjust the value of $\alpha$ by hand to find the exact value of $\alpha$ that satisfies   $(\frac{\partial T}{\partial S})_{\alpha}=0$, which produces $\alpha_c=0.0277925$.} Adapting the same strategy, we also can get the critical value of the Gauss-Bonnet parameter for the case $Q=0.1$, which produces  $\alpha_c=0.01972$. The phase structure for a fixed $Q$ is plotted in  Figure  \ref{fig2}.
\begin{table}
\begin{center}\begin{tabular}{l|c|c|c|c|c|c|}
 \hline
            &$T_{f}$ &     $S_{min}$ &   $S_{max}$     &$A_L$ &    $A_R$ &    $\textrm{relative ~error}$  \\  \hline
$Q=0$        & 0.4163  & 0.126349   & 2.11603 &  0.828192  &    0.828303   &      0.134\%
     \\ \hline
$Q=1$         & 0.4348 & 0.209455 &  2.32746 & 0.920752  &   0.920907      &  0.169\%    \\ \hline
$\alpha=0.01$   &0.4377  & 0.152951 & 2.59566 & 1.06919 &             1.06917  &    0.018\%    \\ \hline
$\alpha=0.2$   &0.4145 &0.184371 & 1.95317 &0.733093 & 0.733169      & 0.010\%     \\ \hline
\end{tabular}
\end{center}
\caption{Check of the equal area law in the $T-S$ plane. Where the relative error is defined by ${A_L-A_R\over A_R}$.}\label{tab1}
\end{table}
As $\alpha$  is fixed, we also can investigate how the charge affects the phase structure of the black hole.  For the case $\alpha=0.01$ and $\alpha=0.02$, the critical charge $Q_c$ are found to be 0.1681103 and 0.094984 respectively. The phase structure for a fixed $\alpha$ is plotted in Figure  \ref{fig4}.

From  Figure  \ref{fig2} and  Figure  \ref{fig4}, we know that these phase structures are similar to that of the Van der Waals phase structure.
That is, the black hole endowed with different charges or Gauss-Bonnet parameters have different phase structures.
As the value of the charge or Gauss-Bonnet parameter is smaller than the corresponding critical value, there is a three special phases region where a small black hole, large black hole and an intermediate black hole coexist.  From data about $T_f$ table \ref{tab1}, one can see $T_f$ will increase with increasing charge $Q$ with fixing $\alpha$($\alpha$ with fixing $Q$) to the critical $T_c$. When $T_f<T_c$, the small black hole will coexist with large black hole.  One can make use of equal area law to determined $T_f$ which is the temperature of coexistence of small black hole and large black hole. While the  temperature increase to $T_c$, the swallow tails will shrink to a critical point and equal area will go to vanishing. The large black hole and small black hole will go to one black hole. This phenomenon will be analogous with the one in Van der Waals fluid below the critical temperature, as the volume decreased a certain pressure is reached in which gas and liquid coexist. In our case, we can map small black and large black to liquid phase and gas phase in fluid system in analogy sense. With increasing the value of the charge or Gauss-Bonnet parameter to the corresponding critical value, the small black hole and the large
black hole will merge into one and squeeze out the unstable phase such that an inflection point emerges. In this situation, the divergence of the heat capacity implies that there is a second order phase transition. For the case that the value  of the charge or Gauss-Bonnet parameter exceeds the corresponding critical value, the black hole is always stable.

{The phase structures can also be observed in the $F-T$ plane, in which $F=M-TS$ is the Helmholtz free energy, where $M$ is black hole mass, with taking the case $Q=0.03,\alpha=0.02$ as an example. From (b) of Figure \ref{fig5}, there is a swallowtail structure, which corresponds to the unstable phase in the top curve in (b) Figure \ref{fig4}.
The transition temperature $T_f=0.4145$ is apparently the value of the horizontal coordinate of the junction between the small black hole and the large
black hole. When the temperature is lower than the transition temperature $T_f$, the free energy of
the small black hole is lowest which means the small hole is stable and dominant. As the temperature is higher than
$T_f$, the free energy of the large black hole is lowest, so the large black hole dominates thereafter.
The non-smoothness of the junction in figure \ref{fig5} indicates that the phase transition is first order.}

{In addition, the critical temperature $T_f$ also satisfy  Maxwell's equal area law
\begin{equation}
\text{ } A_L\equiv\int_{S_1}^{S_3}T(S,Q)dS\text{}=T_{f}(S_3-S_1)\text{}\equiv A_R,\text{  } \label{euqalarea}
 \end{equation}
in which $T(S,Q)$ is the analytical function mentioned above, $S_1$ and  $S_3$ are the smallest and largest roots of the equation $T(S,Q)=T_{f}$.}
For different $Q$ and $\alpha$, the results are listed in  Table \ref{tab1}. From this table, we find $A_L$ equals to $A_R$ within our numerical accuracy. So the equal area law still holds in the $T-S$ plane.

For the second order phase transition in Figure  \ref{fig2} and  Figure  \ref{fig4}, we know that near the critical temperature $T_c$, there is always a relation \cite{Johnson}
{ \begin{eqnarray}
\text{ }\log\mid T-T_c\mid \text{}\text{}=3 \log\mid S- S_c\mid +\text{constant} \text{ }\text{ }   \label{cc2},
 \end{eqnarray}
 in which $S_c$ is the critical entropy corresponding the critical temperature.
 With the definition of the heat capacity
 \begin{equation}
\text{ } C_{Q}=T\frac{\partial S}{\partial T}\Big|_Q \text{ }\label{capacity}.
 \end{equation}
 One can get further $C_Q\sim(T-T_c)^{-2/3}$, namely the critical exponent is $-2/3$, which is the same as the one \cite{Banerjee:2011cz} from the mean field theory.} Next, we will check whether there is a similar relation as  (\ref{cc2}) to check the critical exponent of the heat capacity in the framework of holography.

\section{Phase structure of the non-local observables}
Having understood the phase structure of the black hole from
the viewpoint of thermodynamics, we will employ the non-local observables such as holographic entanglement entropy, Wilson loop, and two point correlation function to probe the phase structure.
The main motivation is to check whether the non-local observables exhibit the similar phase structure
as that of the thermal entropy.
\subsection{Phase structure   probed by holographic entanglement entropy}

The  holographic entanglement entropy in the Gauss-Bonnet gravity can be proposed as \cite{deBoer:2011wk,Hung:2011xb,Fursaev:2006ih, Ogawa:2011fw,Myers:2010xs,Dong:2013qoa,Camps:2013zua,Miao:2014nxa}.
\begin{eqnarray}\label{No}
S_A = \frac{2 \pi}{\ell_p^3} \int_M d^3x\sqrt{h}\,\left[ 1+ \alpha
L^{2} \mathcal{R} \right]+\frac{4\pi}{\ell_p^{3}} \int_{\partial M}
d^{2}x\sqrt{h} \alpha L^{2} \mathcal{K}. \label{GBEE4}
\end{eqnarray}
The first integral in ($\ref{GBEE4}$) is evaluated on the bulk surface $M$, the second one is on boundary $\partial M$, which is the boundary of $ M$ regularized at $r=r_0$, $\mathcal{R}$ is the Ricci scalar for the intrinsic metric
of $M$, $\mathcal{K}$ is the trace of the extrinsic
curvature of the boundary of $M$ and $h$ is the determinant of the
induced metric on $M$. The second term in the first integral ($\ref{GBEE4}$) is
present due to higher derivative gravity appeared in the background. The
minimal value of the functional ($\ref{GBEE4}$) would give the
entanglement entropy of the subsystem $A$.

For our background,
the entangling
surface is parameterized as a constant $\theta$ hypersurface $\theta=\theta_0$ with coordinates
$0\leq \phi \leq \pi, 0\leq \psi \leq 2\pi.$ In this case, based on  (\ref{GBEE4}), we can get the equation of motion of $r(\theta)$
\begin{eqnarray}
&& \pi ^2 r(\theta )(r'(\theta )^2 (\sin (\theta ) r(\theta )^2 f'(r(\theta ))-2 \cos (\theta ) r'(\theta ))-2 r(\theta ) f(r(\theta )) (r(\theta ) (\sin (\theta ) r''(\theta )
\nonumber\\
&&-3+\cos (\theta ) r'(\theta )) \sin (\theta ) r'(\theta )^2)+4 \sin (\theta ) r(\theta )^3 f(r(\theta ))^2)=0,
\end{eqnarray}
in which ${r}^{\prime}=dr/ d\theta$.
To solve this equation, we will resort to the following boundary conditions
\begin{eqnarray}
{r}^{\prime}(0)=0, r(0)= r_0. \label{bon}
\end{eqnarray}

\begin{figure}[h!]
\centering
\subfigure[$Q=0$]{
\includegraphics[scale=0.6]{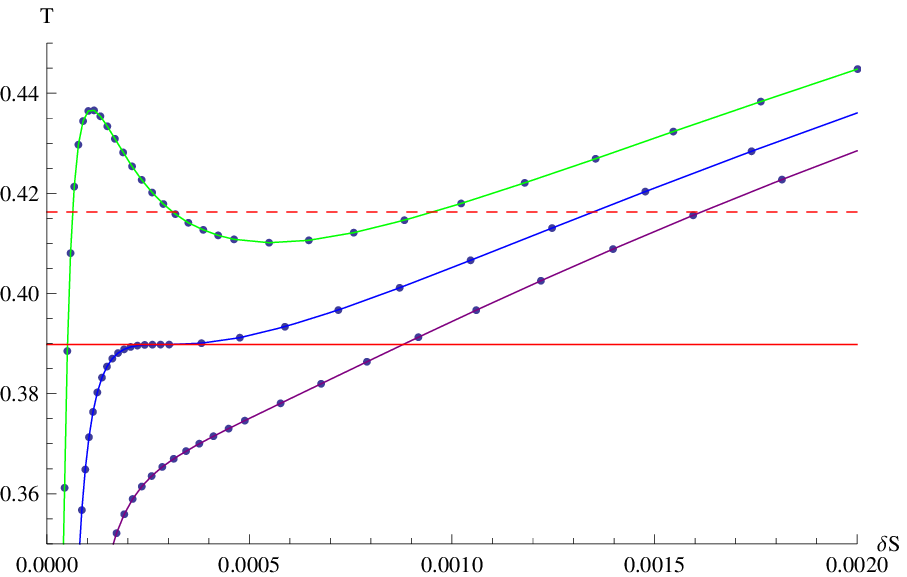}
}
\subfigure[$Q=0.1$]{
\includegraphics[scale=0.6]{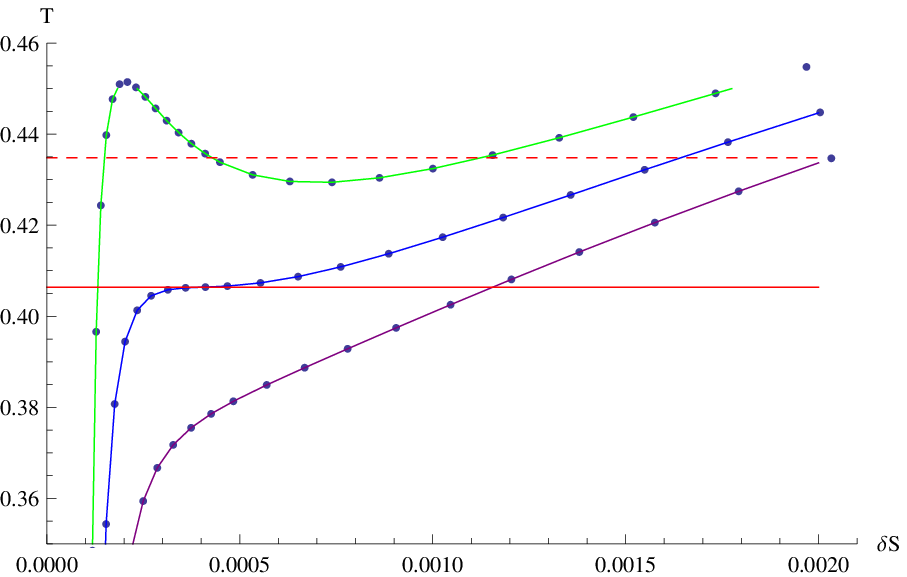}
}
\caption{\label{fig6}\small Relations between  holographic entanglement entropy and temperature  for different $\alpha$   at a fixed $Q$. In (a), curves from top to down correspond to $\alpha=0.02,0.0277925, 0.035$, and  in (b) they correspond to $\alpha=0.01,0.01972, 0.03$ respectively. The red   dashed  line and solid line correspond to the first order  phase transition temperature $T_f$ and second order phase transition temperature $T_c$. }
\end{figure}

\begin{figure}[h!]
\centering
\subfigure[$\alpha=0.01$]{
\includegraphics[scale=0.6]{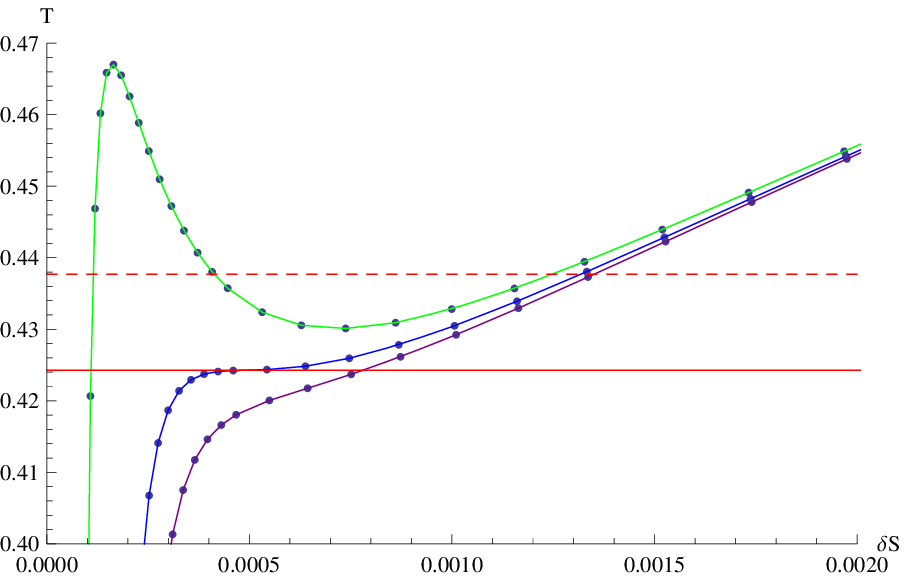}
}
\subfigure[$\alpha=0.02$]{
\includegraphics[scale=0.6]{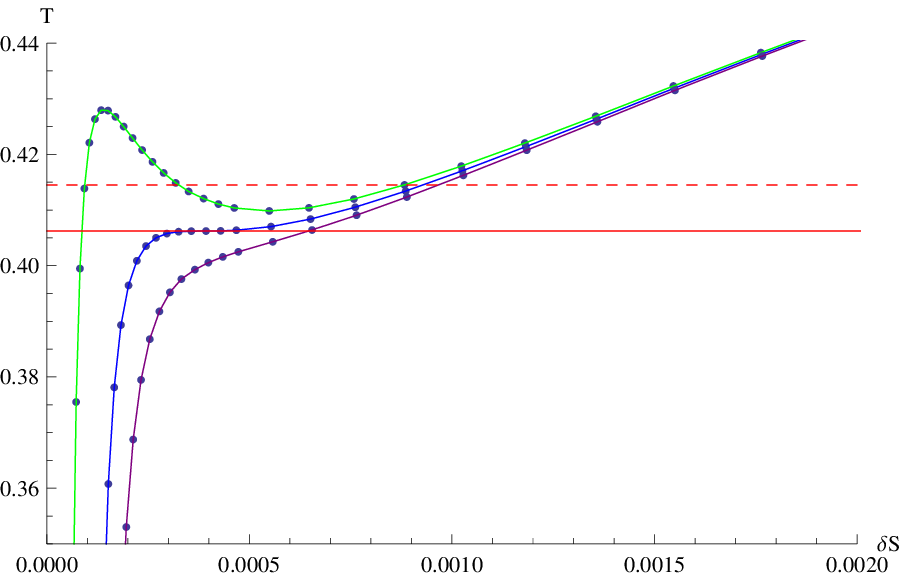}
}
\caption{\label{fig7}\small Relations between  holographic entanglement entropy and temperature  for different $Q$ at a fixed $\alpha$. In (a), curves from top to down correspond to $Q=0.08,0.1681103, 0.2$, and  in (b) they correspond to $Q=0.03,0.094984, 0.13$ respectively. The red   dashed  line and solid line correspond to the first order  phase transition temperature $T_f$ and second order phase transition temperature $T_c$. }
\end{figure}

In addition, to avoid the entanglement entropy to be contaminated by the surface that wraps
the horizon, we will choose a small region as $A$ as in \cite{Johnson}. In this paper, we  will choose  $\theta_0=0.2$. Note that for a fixed $\theta_0$, the entanglement entropy  is divergent, so it should be regularized by subtracting off the entanglement entropy in pure AdS with the same  boundary region, denoted by $S_0$.  To achieve this, we are required to set a UV cutoff, which is chosen to be $r_0= r(0.199)$.
The regularized  entanglement entropy is labeled as $\delta S\equiv S_A-S_0$.

With these assumption, we can plot the phase structure of entanglement entropy for a fixed charge $Q$ or a fixed Gauss-Bonnet parameter $\alpha$, which are shown in Figure \ref{fig6} and  Figure  \ref{fig7}. It is obvious that  Figure \ref{fig6} and  Figure  \ref{fig7} reassemble as Figure \ref{fig2} and  Figure  \ref{fig4} respectively. Especially, the first order phase transition temperature $T_f$ and second order phase transition temperature $T_c$ are exactly the same as that in the $T-S$ plane.  We
will employ the equal area law to locate the first order phase transition temperature, and  critical exponent of the analogous heat capacity to locate  the second order phase transition temperature.

Similar to (\ref{euqalarea}), the equal area law in $T-\delta S$ plane can be defined as
\begin{equation}
\text{}A_L\equiv\int_{\delta S_{min}}^{\delta S_{max}}T(\delta S,Q)\text{}d\delta S=T_f\text{}(\delta S_{max}-\delta
S_{min})\equiv A_R \label{eeuqalarea},
 \end{equation}
in which  $T(\delta S)$  is an interpolating function obtained from the numeric data, $T_{f} $ is the phase transition temperature, and $\delta S_{min}$,  $\delta S_{max}$ are the smallest and largest values of the  equation $T(\delta S)=T_{f}$.
For different $Q$ and $\alpha$, the calculated results are listed in Table  \ref{tab2}. From this table, we can see that for the unstable region of the first order phase transition in the $T-\delta S$ plane, the equal area law holds within our numeric accuracy.

\begin{table}
\begin{center}\begin{tabular}{l|c|c|c|c|c|c|}
 \hline
            &$T_{f}$ &     $\delta S_{min}$ &   $\delta S_{max}$     &$A_L$ &    $A_R$ &    $\textrm{relative ~error}$  \\  \hline
$Q=0$        & 0.4163  &$ 0.00006408$   &$0.0009523$ & $0.00036975$ &   $0.00036984$ &      0.02466\%
     \\ \hline
$Q=0.1$        & 0.4348  &$0.0001495$   &$0.001124$ & $ 0.00042361 $ &   $0.00042367 $ &      0.01496\%   \\ \hline
$\alpha=0.01$   & 0.4377  &$0.0001149$   &$0.0012467$ & $ 0.00049540$ &   $0.00049571$ &      0.06335\%   \\ \hline
$\alpha=0.02$   & 0.4145  &$ 0.00009349$   &$0.0008822$ & $ 0.00032692 $ &   $0.00032696$ &      0.01147\%    \\ \hline
\end{tabular}
\end{center}
\caption{Check of the equal area law in the $T-\delta S$ plane. Where the relative error is defined by ${A_L-A_R\over A_R}$.}\label{tab2}
\end{table}

\begin{figure}[h!]
\centering
\subfigure[$Q=0,\alpha=0.0277925$]{
\includegraphics[scale=0.6]{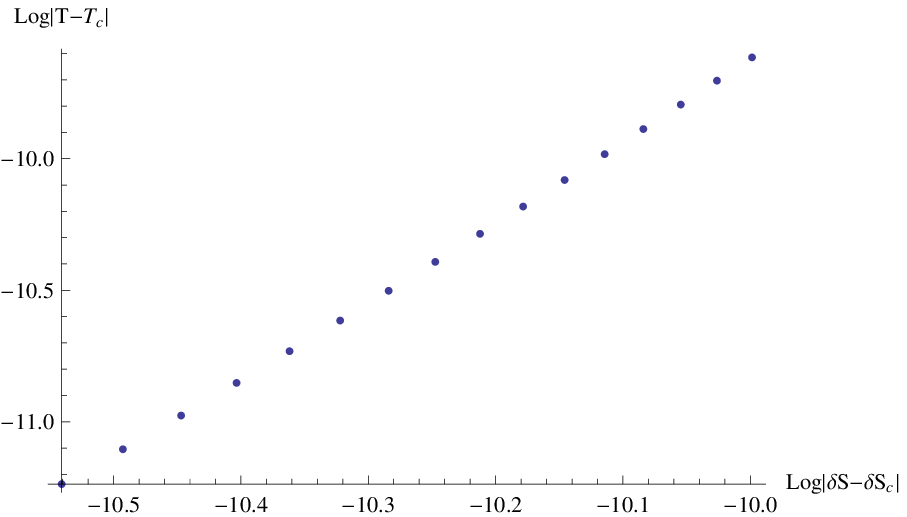}
}
\subfigure[$Q=0.1,\alpha=0.01972$]{
\includegraphics[scale=0.6]{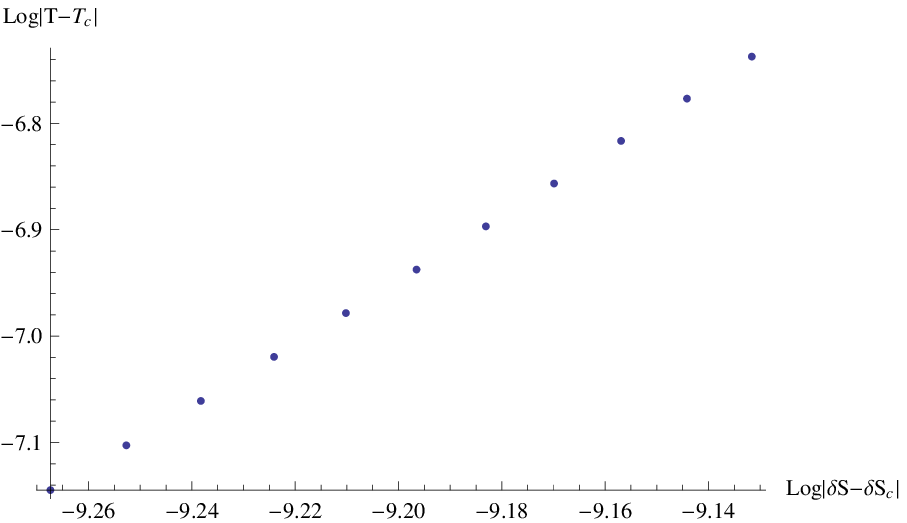}
}
\subfigure[$Q=0.1681103,\alpha=0.01$]{
\includegraphics[scale=0.6]{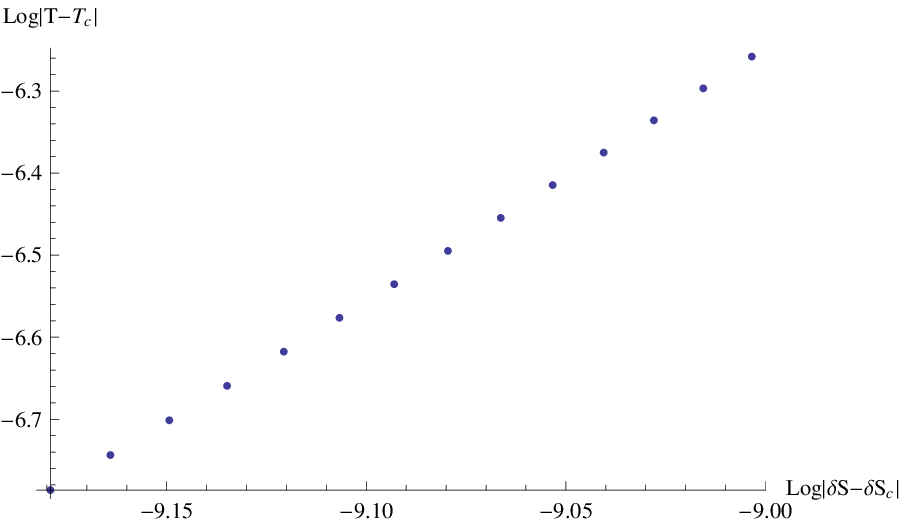}
}
\subfigure[$Q=0.094984,\alpha=0.02$]{
\includegraphics[scale=0.6]{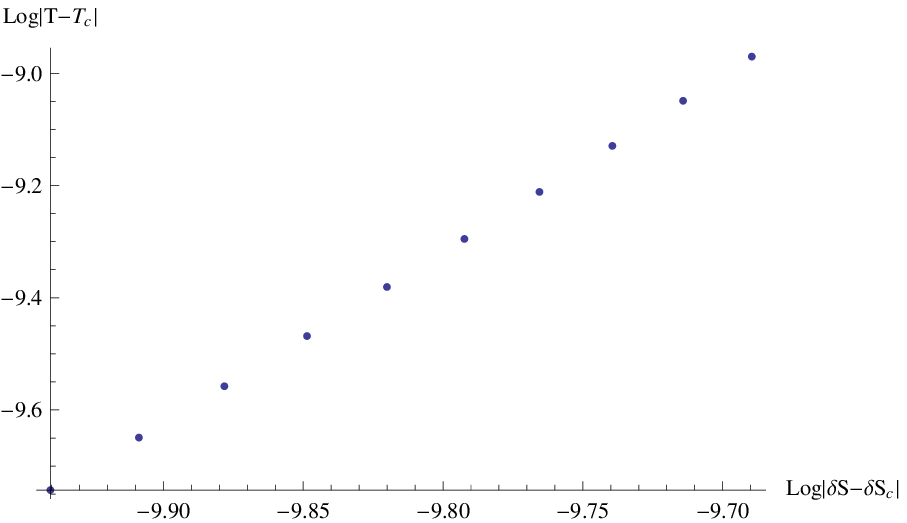}
}
\caption{\label{fig8}\small Relations between $\log\mid T-T_c\mid$ and  $\log\mid\delta S-\delta S_c\mid $ for different $Q$ and $\alpha$ }
\end{figure}

In order to investigate the critical exponent of the second order phase transition in the $T-\delta S$  plane, we  define an analogous heat capacity
\begin{eqnarray}
C=T\frac{\partial \delta S}{\partial T}. \label{cheat2}
 \end{eqnarray}
Provided a similar relation as showed  in (\ref{cc2}) is satisfied, one can get the critical exponent of the analogous heat capacity immediately.
So next, we are interested in the logarithm of the quantities $T-T_c$, $\delta S-\delta S_c$, in which $T_c$ is the second order phase transition temperature, and
 $\delta S_c$ is  obtained numerically by the equation $T(\delta S)=T_c$.
 The relation between $ \log\mid T -T_c\mid$ and $\log\mid\delta S-\delta S_c\mid  $ for different $Q$  and $\alpha$ are shown in Figure \ref{fig8}. By data fitting, the
straight lines  in  Figure \ref{fig8} can be expressed as
\begin{equation}
\log\mid T-T_c\mid=\begin{cases}
20.3652\, +3.00026 \log\mid\delta S-\delta S_c\mid,&$for$ ~Q=0,\alpha=0.0277925,\\
20.668\, +3.0015 \log\mid\delta S-\delta S_c\mid, & $for$~   Q=0.1,\alpha=0.01972,\\
20.817\, +3.00773 \log\mid\delta S-\delta S_c\mid,&$for$~Q=0.1681103,\alpha=0.01,\\
20.8815 +3.08132  \log\mid\delta S-\delta S_c\mid, & $for$~Q=0.094984,\alpha=0.02.\\
\end{cases}
\end{equation}
It is obvious that for all the lines, the slope is about 3, which resembles as that in (\ref{cc2}). That is, the critical exponent of the analogous heat capacity  in $T-\delta S$ plane is the same as that in the $T-S$ plane, which once reinforce the conclusion that the phase structure of the entanglement entropy is the same as that of the thermal entropy.

\subsection{Phase structure   probed by Wilson loop}

\begin{figure}[h!]
\centering
\subfigure[$Q=0$]{
\includegraphics[scale=0.6]{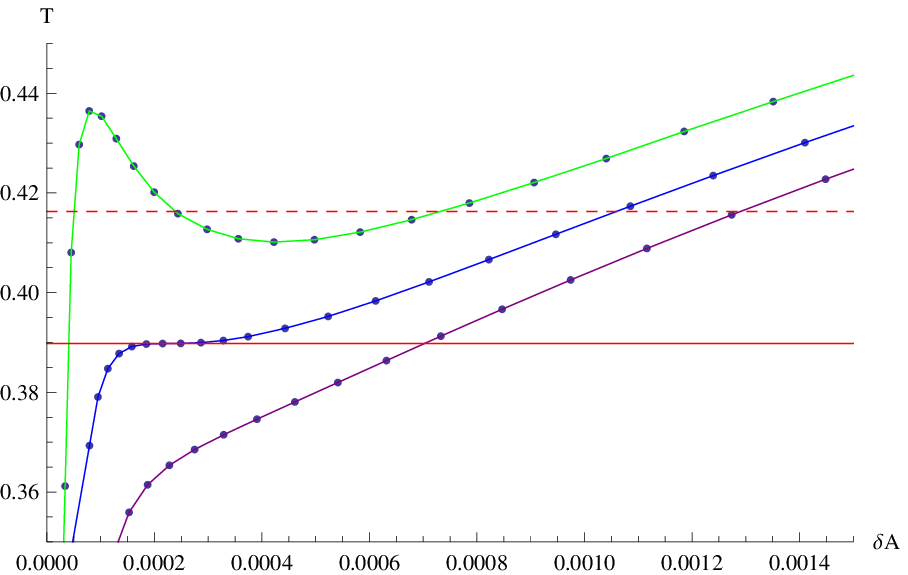}
}
\subfigure[$Q=0.1$]{
\includegraphics[scale=0.6]{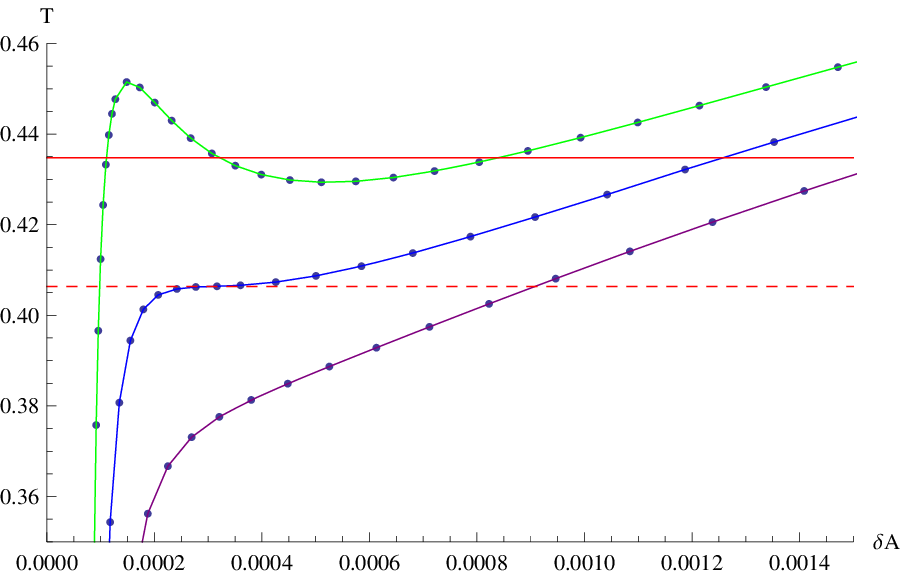}
}
\caption{\small \label{fig9} Relations between minimal area surface and temperature  for different $\alpha$   at a fixed $Q$. In (a), curves from top to down correspond to $\alpha=0.02,0.0277925, 0.035$, and  in (b) they correspond to $\alpha=0.01,0.01972, 0.03$ respectively. The red   dashed  line and solid line correspond to the first order  phase transition temperature $T_f$ and second order phase transition temperature $T_c$. }
\end{figure}

\begin{figure}[h!]
\centering
\subfigure[$\alpha=0.01$]{
\includegraphics[scale=0.6]{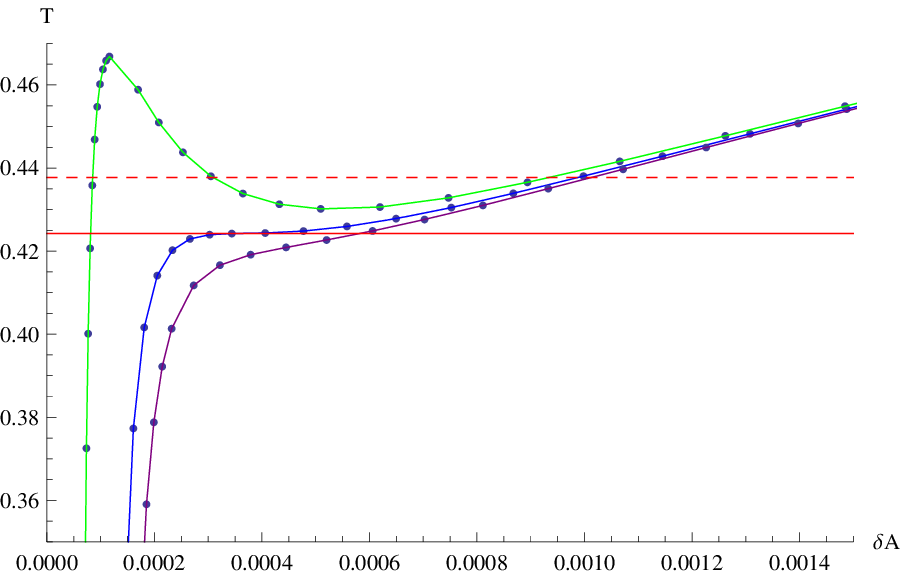}
}
\subfigure[$\alpha =0.02$]{
\includegraphics[scale=0.6]{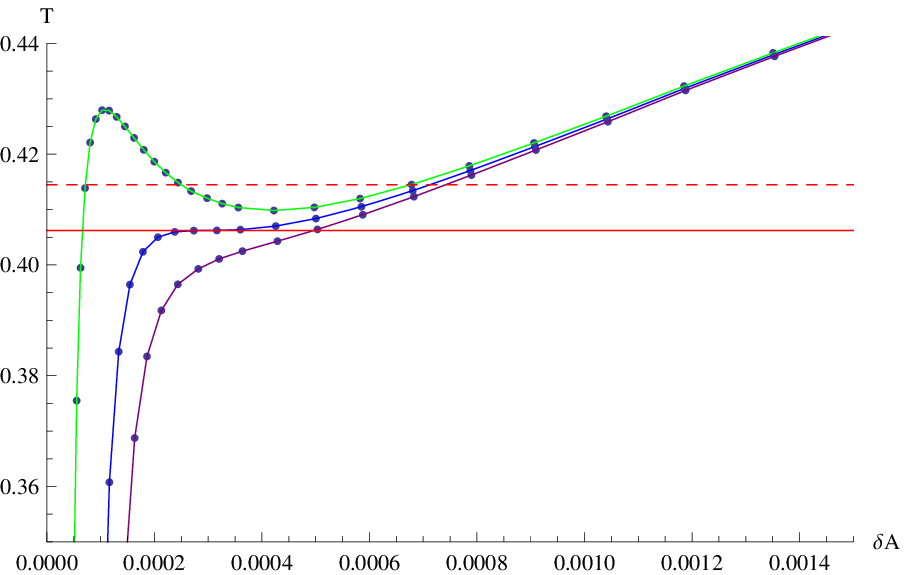}
}
\caption{\small \label{fig10}  Relations between minimal area surface and temperature  for different $Q$ at a fixed $\alpha$. In (a), curves from top to down correspond to $Q=0.08,0.1681103, 0.2$, and  in (b) they correspond to $Q=0.03,0.094984, 0.13$ respectively. The red   dashed  line and solid line correspond to the first order  phase transition temperature $T_f$ and second order phase transition temperature $T_c$. }
\end{figure}

In this subsection, we will employ the Wilson loop to probe the phase structure of the Gauss-Bonnet-AdS black hole.
According to
the AdS/CFT correspondence, the expectation value of the Wilson loop is related to the string
partition function
\begin{equation} \label{area}
\text{}\langle W(C)\rangle = \text{}\int D \Sigma e^{-A(\Sigma_0)}\text{},
\end{equation}
in which
$C$ is the closed contour, $\Sigma_0$ is the string world sheet which extends in the bulk with the boundary condition $\partial \Sigma_0= C$, and $ A(\Sigma_0)$ corresponds to the Nambu-Goto action for the string. In the strongly coupled limit, we
can simplify the computation by making a saddle point approximation and evaluating the minimal
area surface of the classical string with the same boundary condition  $\partial \Sigma_0= C$,
which leads to \cite{Maldacena3}

\begin{equation} \label{area}
\text{}\langle W(C)\rangle \text{}\text{}\text{}\approx e^{-A(\Sigma)},
\end{equation}
where $\Sigma$ represents the minimal area surface.
Next we choose the line with $\phi=\frac{\pi}{2}$ and $\theta=\theta_0$ as our loop. Then we can employ  $(\theta, \psi)$  to parameterize the minimal area surface, which is invariant under the $\psi$-direction by our rotational symmetry.
 Thus the
corresponding minimal area surface   can be expressed as
\begin{eqnarray}
\text{}A= 2\pi \text{}\int_0 ^{\theta_0} r \sin \text{}\theta \sqrt{\frac{{r}^{\prime 2}}{f(r)}+r^2} \text{}  d\theta,
 \end{eqnarray}
Similar to the case of entanglement entropy, we will also use the boundary condition in (\ref{bon}) to solve  $r(\theta)$ with the choice
 $\theta_0= 0.2$.
We label the regularized minimal area surface as $\delta A\equiv A-A_0$, where  $A_0$ is the minimal area in pure AdS with the same  boundary region.
  We plot the relation between  $\delta A$ and $T$  for different  $Q$ and $\alpha$ in Figure \ref{fig9} and  Figure \ref{fig10}. Comparing  Figure \ref{fig9} and  Figure \ref{fig10} with  Figure \ref{fig2} and  Figure \ref{fig4}, we find they are the same nearly besides the scale of the horizonal coordinate.  The result tells us that the similar phase structure also shows up for the minimal surface area, which is the same as that of the entanglement entropy.

\begin{table}
\begin{center}\begin{tabular}{l|c|c|c|c|c|c|}
 \hline
            &$T_{f}$ &     $\delta A_{min}$ &   $\delta A_{max}$     &$A_L$ &    $A_R$ &    $\textrm{relative ~error}$  \\  \hline
$Q=0$        & 0.4163  &$ 0.00007374$   &$0.0007352$ & $ 0.0002742 $ &   $0.0002754$ &      0.4048\%
     \\ \hline
$Q=0.1$        & 0.4348  &$ 0.0001110$   &$0.0008404$ & $ 0.0003172 $ &   $0.0003171 $ &      0.008852\%   \\ \hline
$\alpha=0.01$   & 0.4377  &$ 0.00006426$   &$0.0009327$ & $ 0.0003785$ &   $0.0003801$ &      0.4285\%   \\ \hline
$\alpha=0.02$   & 0.4145  &$ 0.00007167$   &$0.0006778$ & $ 0.0002513 $ &   $0.0002512$ &      0.01042\%    \\ \hline
\end{tabular}
\end{center}
\caption{Check of the equal area law in the $T-\delta A$ plane. Where the relative error is defined by ${A_L-A_R\over A_R}$.}\label{tab4}
\end{table}

\begin{figure}[h!]
\centering
\subfigure[$Q=0,\alpha=0.0277925$]{
\includegraphics[scale=0.6]{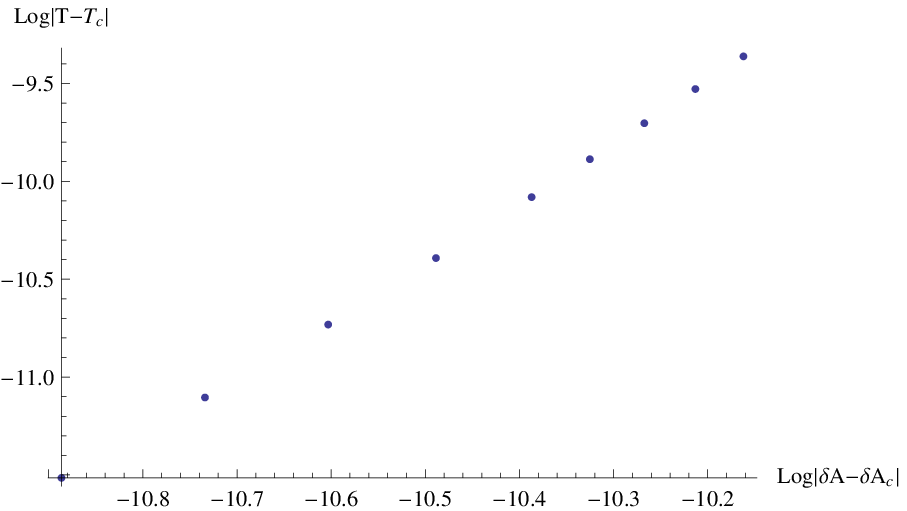}
}
\subfigure[$Q=0.1,\alpha=0.01972$]{
\includegraphics[scale=0.6]{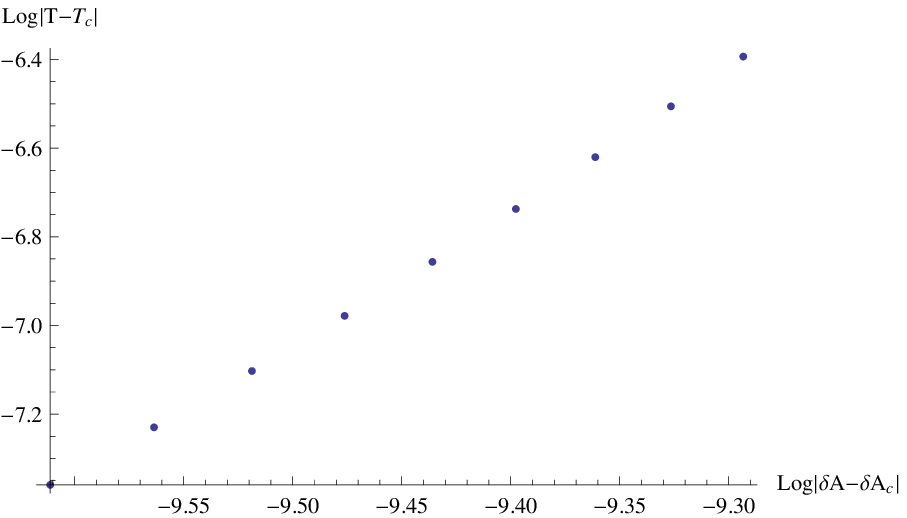}
}
\subfigure[$\alpha=0.01,Q=0.1681103$]{
\includegraphics[scale=0.6]{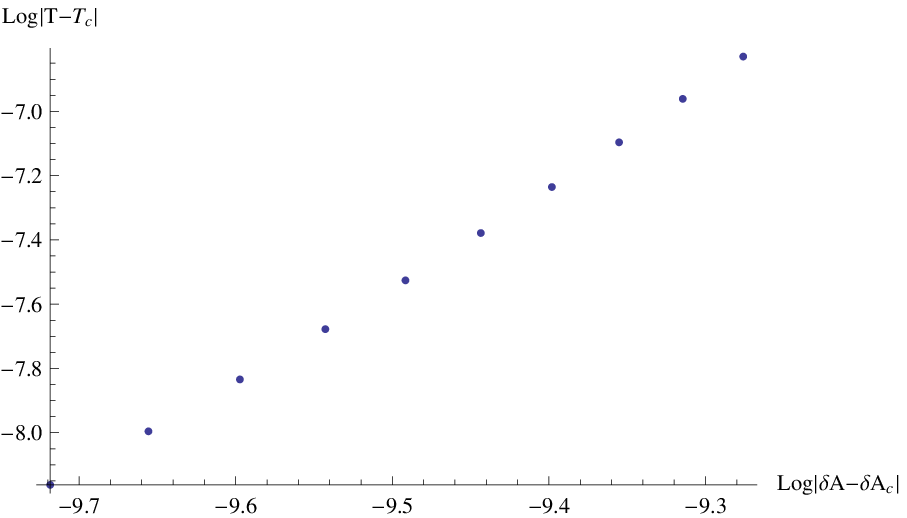}
}
\subfigure[$\alpha=0.02 ,Q=0.094984$]{
\includegraphics[scale=0.6]{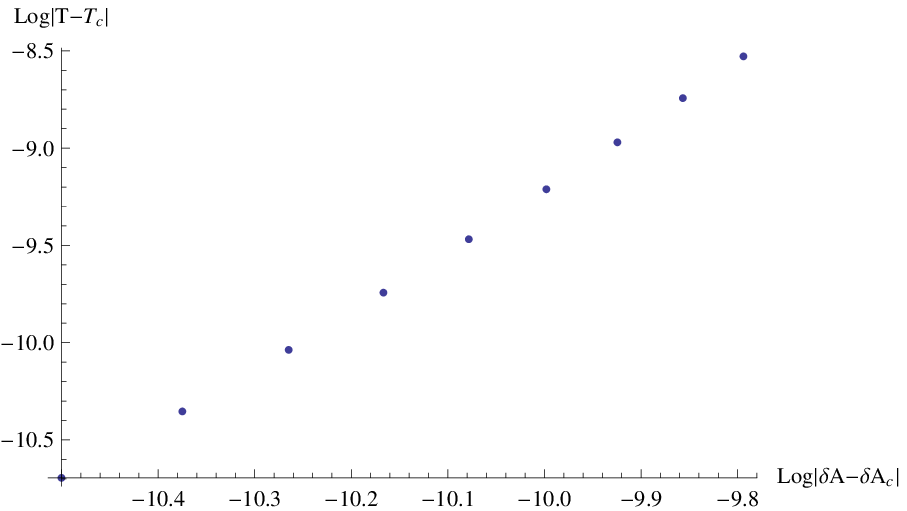}
}
\caption{\small \label{fig11}   Relations between $\log\mid T-T_c\mid$ and  $\log\mid\delta A-\delta A_c\mid $ for different $Q$ and $\alpha$ }
\end{figure}
It is also necessary to check  the equal area law for the first order phase transition and   critical exponent of the analogous heat capacity for the second order phase transition in the $T-\delta A$ plane.  The equal area law in  this case can be defined as
\begin{equation}
\text{}A_L\equiv\int_{\delta A_{min}}^{\delta A_{max}}T(\delta A,Q)d \delta A \text{}= \text{}T_f(\delta A_{max}-\delta
A_{min})\equiv A_R \label{eeuqalarea},
 \end{equation}
in which  $\delta A_{min}$,  $\delta A_{max}$ are the smallest and largest values of the  equation $T(\delta A)=T_{f}$, where  $T(\delta A)$ is an interpolating function obtained by data fitting.
For different $Q$ and $\alpha$, the calculated results are listed in Table  \ref{tab4}. Obviously, as that in the $T-\delta S$ plane, the equal area law also holds  within our numeric accuracy, which implies that the minimal area surface owns the same first order phase transition as that of the thermal entropy.

For the second order phase transition, we are interested in the logarithm of the quantities $T-T_c$, $\delta A-\delta A_c$, in which $T_c$ is the second order phase transition temperature, and
 $\delta A_c$ is  obtained numerically by the equation $T(\delta A)=T_c$.
 The relation between $ \log\mid T -T_c\mid$ and $\log\mid\delta A-\delta A_c\mid  $ for different $Q$  and $\alpha$ are shown in Figure \ref{fig11}. The straight line in each subgraph can be fitted as
\begin{equation}
\log\mid T-T_c\mid=\begin{cases}
20.9318 + 2.98393 \log\mid\delta A-\delta A_c\mid, &$for$ ~Q=0,\alpha=0.0277925,\\
21.843 + 3.04022 \log\mid\delta A-\delta A_c\mid, &$for$ ~ Q=0.1,\alpha=0.01972,\\
21.1332 + 3.0174 \log\mid\delta A-\delta A_c\mid,&$for$~Q=0.1681103,\alpha=0.01,\\
21.5862 + 3.07862  \log\mid\delta A-\delta A_c\mid, & $for$~Q=0.094984,\alpha=0.02.\\
\end{cases}
\end{equation}
Similar to that of the entanglement entropy, the slope of the fitted straight line is also about 3, which implies that the critical exponent of the
analogous heat capacity  is -2/3 in the $T-\delta A$ plane. This result is consistent with that of the thermal entropy too, which reminds that the minimal area surface exhibits the same second order phase transition as that of the thermal entropy.

\subsection{Phase structure   probed by two point correlation function}

\begin{figure}[h!]
\centering
\subfigure[$Q=0,\alpha=0.0277925$]{
\includegraphics[scale=0.6]{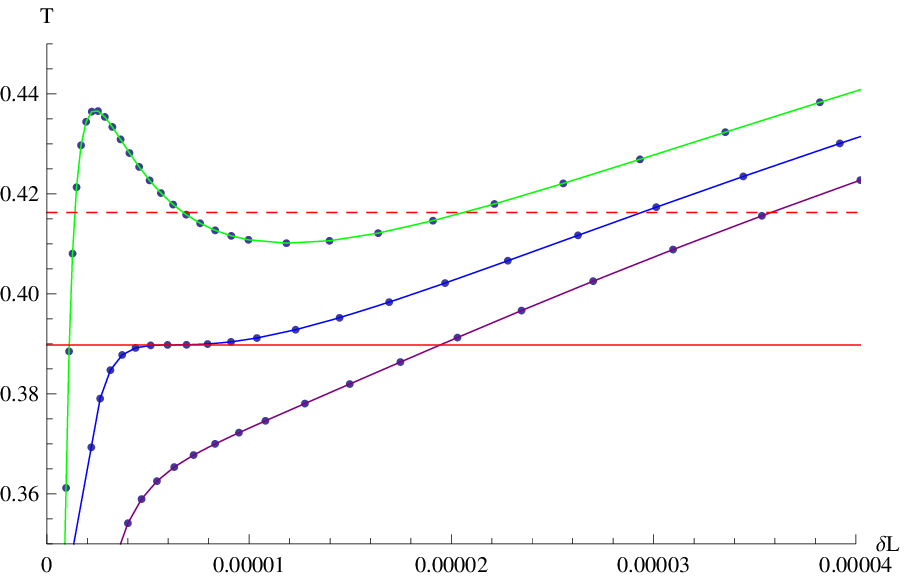}
}
\subfigure[$Q=0.1,\alpha=0.01972$]{
\includegraphics[scale=0.6]{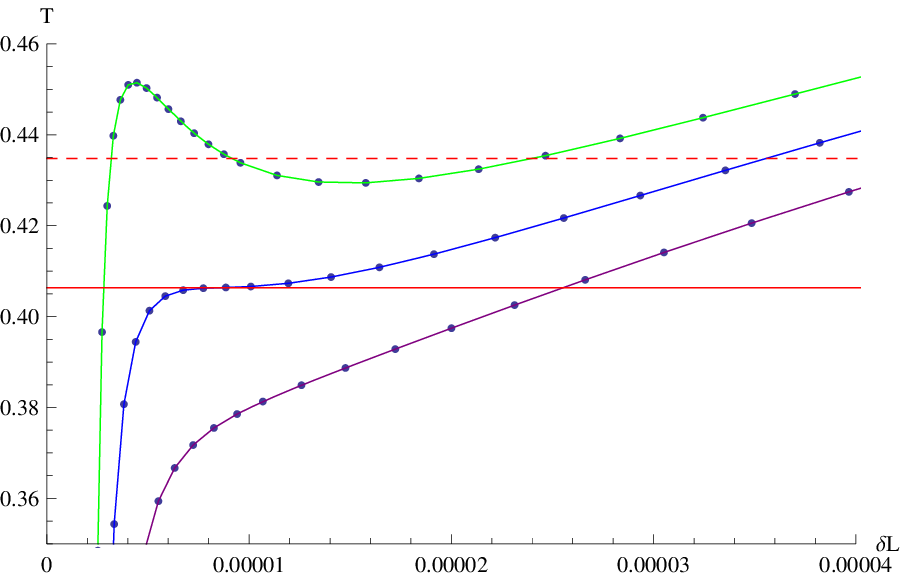}
}
\caption{\small \label{fig12}Relations between geodesic length and temperature  for different $\alpha$   at a fixed $Q$. In (a), curves from top to down correspond to $\alpha=0.02,0.0277925, 0.035$, and  in (b) they correspond to $\alpha=0.01,0.01972, 0.03$ respectively. The red   dashed  line and solid line correspond to the first order  phase transition temperature $T_f$ and second order phase transition temperature $T_c$.  }
\end{figure}

\begin{figure}[h!]
\centering
\subfigure[$\alpha=0.01$]{
\includegraphics[scale=0.6]{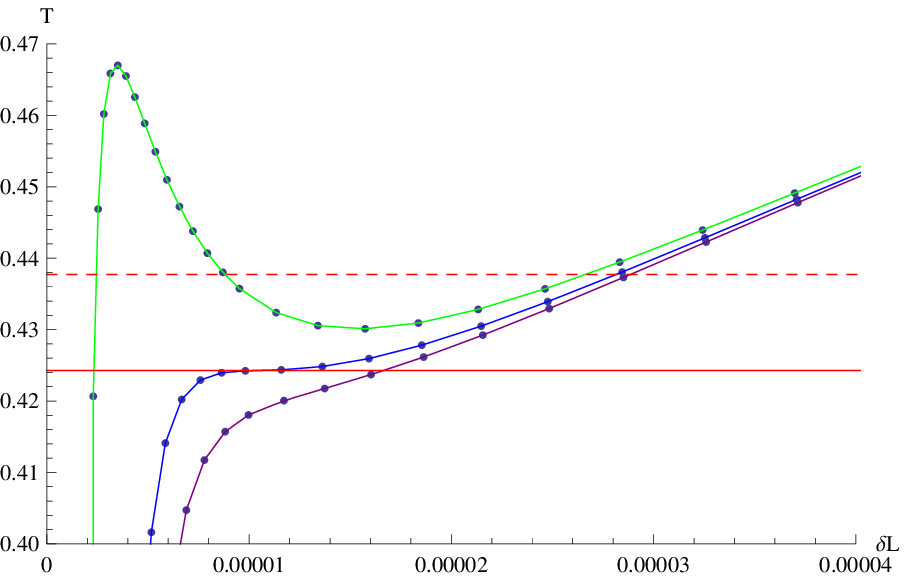}
}
\subfigure[$\alpha=0.02$]{
\includegraphics[scale=0.6]{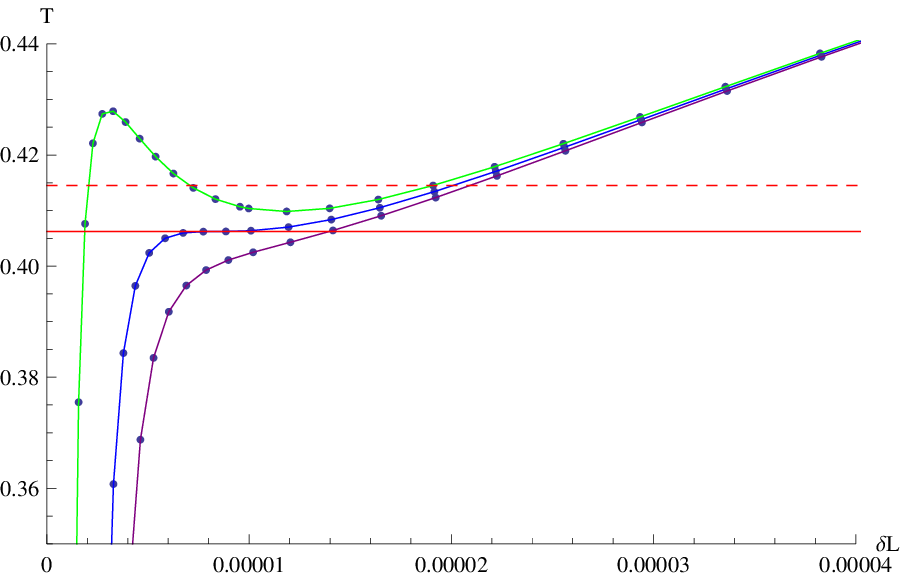}
}
\caption{\small \label{fig13}  Relations between geodesic length and temperature  for different $Q$ at a fixed $\alpha$. In (a), curves from top to down correspond to $Q=0.08,0.1681103, 0.2$, and  in (b) they correspond to $Q=0.03,0.094984, 0.13$ respectively. The red   dashed  line and solid line correspond to the first order  phase transition temperature $T_f$ and second order phase transition temperature $T_c$. }
\end{figure}

In this subsection, we would like to study a scalar operator $\cal{O}$ with large conformal dimension $\Delta$ in the  dual field theory. Due to the saddle point approximation, the equal time two point correlation function can be written as following with following \cite{Balasubramanian61}
 \begin{equation}\label{llll4}
\text{ }\big\langle {\cal{O}} (t,x_i) \text{}\text{}{\cal{O}}(t, x_j)\big\rangle  \approx
e^{-\Delta {\overline{L}}} ,
\end{equation}
in which $\overline{L}$ is the  length of the bulk geodesic between the points $(t,
x_i)$ and $(t, x_j)$ on the AdS boundary.
In our gravity model,
  we can simply choose $(\phi=\frac{\pi}{2},\theta=0, \psi=0)$ and $(\phi=\frac{\pi}{2},\theta=\theta_0,\psi=\pi)$ as the two boundary points. Then with  $\theta$ to
    parameterize the trajectory, the proper length  is given by
\begin{eqnarray}\label{rl}
\overline{L}=\int_0 ^{\theta_0}\sqrt{\frac{{r}^{\prime 2}}{f(r)}+r^2} d\theta,
 \end{eqnarray}
  With the boundary condition in (\ref{bon}), we can
get the numeric result of $r(\theta)$ and further get $\overline{L}$ by substituting $r(\theta)$ into  (\ref{rl}).  Similarly, we label the regularized geodesic length as $\delta L\equiv \overline{L}-\overline{L}_0$, where  $\overline{L}_0$ is the geodesic length in pure AdS with the same  boundary region.
The relation between  $\delta L$ and $T$  for different  $Q$ and $\alpha$ are shown in Figure \ref{fig12} and  Figure \ref{fig13}. It is obvious that Figure \ref{fig12} and  Figure \ref{fig13} resemble as  Figure \ref{fig2} and  Figure \ref{fig4} respectively besides the scale of horizontal coordinate, which implies that the geodesic length owns the same phase structure as that of the thermal entropy. Especially they  have the same first order phase transition temperature and second order phase transition temperature, which will be checked next by investigating the equal area law for the first order phase transition and critical exponent for the second order phase transition.

\begin{table}
\begin{center}\begin{tabular}{l|c|c|c|c|c|c|}
 \hline
            &$T_{f}$ &     $\delta L_{min}$ &   $\delta L_{max}$     &$A_L$ &    $A_R$ &    $\textrm{relative ~error}$  \\  \hline
$Q=0$        & 0.4163  &$ 2.1207\times10^-6$   &$0.00002072$ & $ 7.7104\times10^-6 $ &   $7.7418\times
10^-6 $ &      0.4058\%
     \\ \hline
$Q=0.1$        & 0.4348  &$ 3.2756\times10^-6$   &$0.00002398$ & $ 9.0415\times10^-6 $ &   $9.0402\times10^-6 $ &      0.0147\%   \\ \hline
$\alpha=0.01$   & 0.4377  &$ 2.0984\times10^-6$   &$0.00002661$ & $ 0.00001051 $ &   $0.000010727$ &      1.9899\%   \\ \hline
$\alpha=0.02$   & 0.4145  &$ 1.3642\times10^-6$   &$0.00001894$ & $ 7.3222\times10^-6 $ &   $7.2869\times
10^-6 $ &      0.4820\%    \\ \hline
\end{tabular}
\end{center}
\caption{Check of the equal area law in the $T-\delta L $ plane. Where the relative error is defined by ${A_L-A_R\over A_R}$.}\label{tab3}
\end{table}

\begin{figure}[h!]
\centering
\subfigure[$Q=0,\alpha=0.0277925$]{
\includegraphics[scale=0.6]{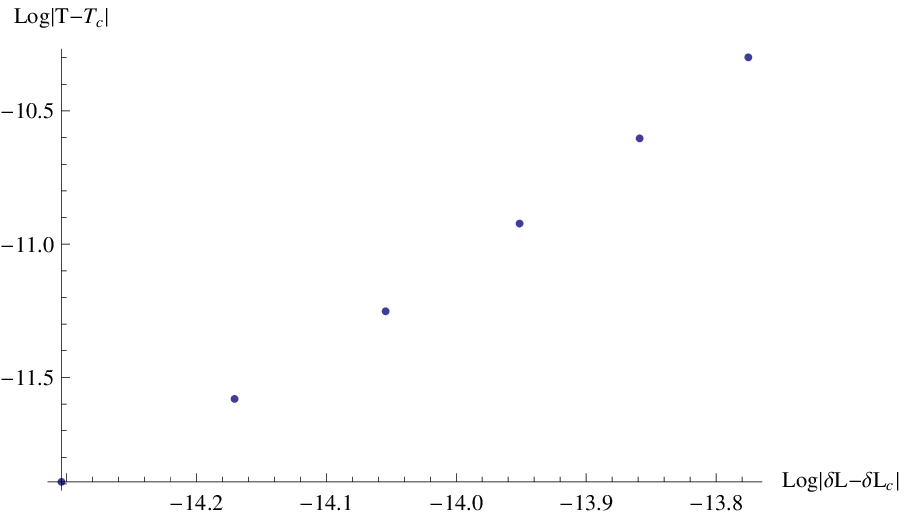}
}
\subfigure[$Q=0.1,\alpha=0.01972$]{
\includegraphics[scale=0.6]{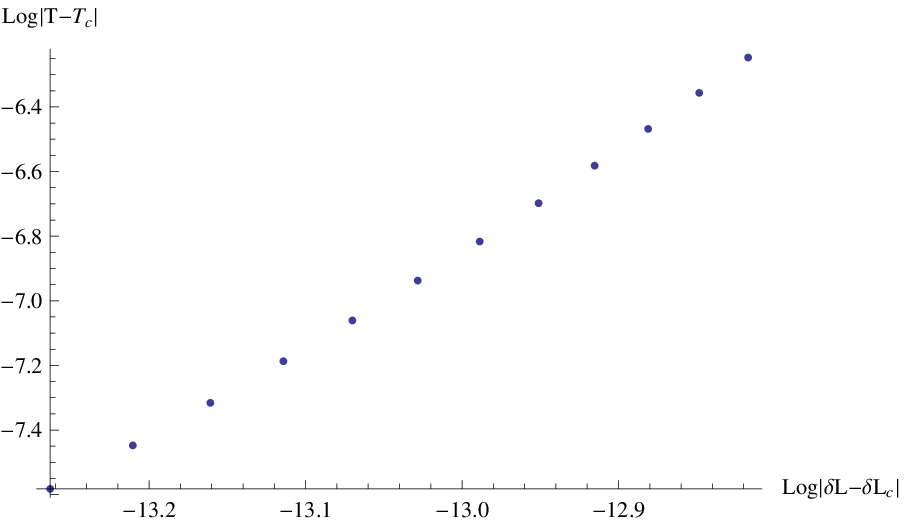}
}
\subfigure[$Q=0.1681103,\alpha=0.01$]{
\includegraphics[scale=0.6]{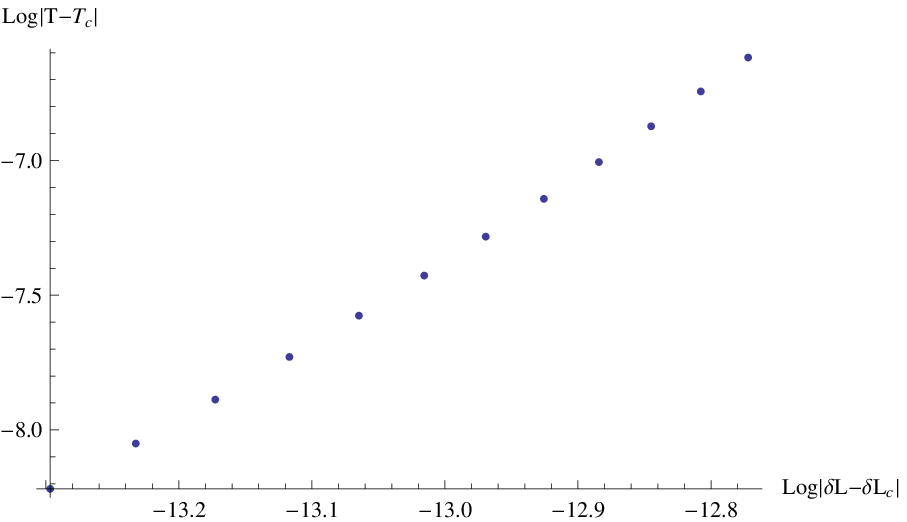}
}
\subfigure[$Q=0.094984,\alpha=0.02$]{
\includegraphics[scale=0.6]{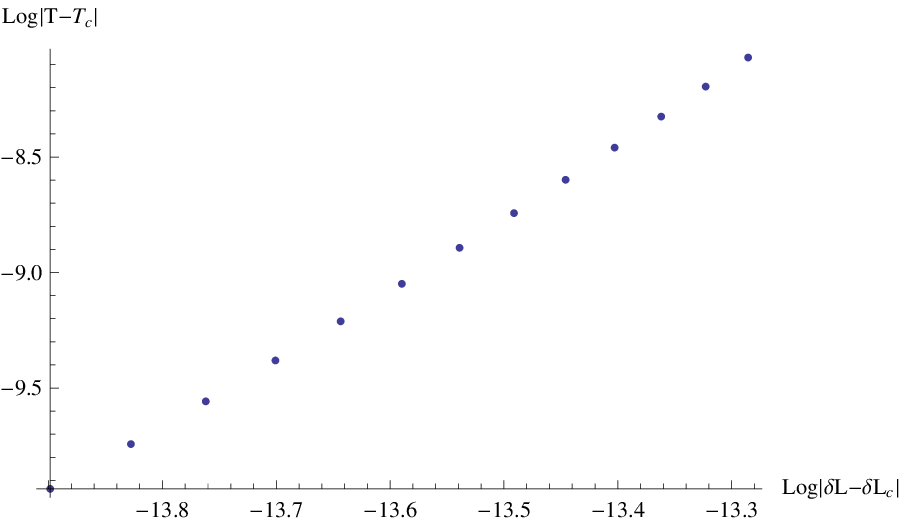}
}
\caption{\small \label{fig14}  Relations between $\log\mid T-T_c\mid$ and  $\log\mid\delta L-\delta L_c\mid $ for different $Q$ and $\alpha$  }
\end{figure}

{In the $T-\delta L$ plane, The equal area law can be defined as
\begin{equation}
\text{ }\text{ }A_L\equiv\text{ }\int_{\delta L_{min}}^{\delta L_{max}}T(\delta L,Q)d \text{ } \delta L=\text{}T_f(\delta L_{max}-\delta
L_{min})\equiv A_R \text{ }\text{ }\label{Leuqalarea},
 \end{equation}
in which  $\delta L_{min}$,  $\delta L_{max}$ are the smallest and largest values of the  equation $T(\delta L)=T_{f}$, where  $T(\delta L)$ is also an interpolating function.
For different $Q$ and $\alpha$, the  results are listed in Table  \ref{tab3}.  We can see that in the $T-\delta L$ plane, the equal area law holds  within a reasonable  numeric accuracy.}

For the second order phase transition, we will investigate the  relation between $ \log\mid T -T_c\mid$ and $\log\mid\delta L-\delta L_c\mid  $ for different $Q$  and $\alpha$, which are shown in Figure  \ref{fig14}. The straight lines in this figure can be fitted respectively as
\begin{equation}
\log\mid T-T_c\mid=\begin{cases}
31.3704 + 3.02888 \log\mid\delta L-\delta L_c\mid,&$for$ ~Q=0,\alpha=0.0277925,\\
32.1586 + 2.99908 \log\mid\delta L-\delta L_c\mid, &$for$ ~ Q=0.1,\alpha=0.01972,\\
32.4443 + 3.06135 \log\mid\delta L-\delta L_c\mid,&$for$~Q=0.1681103,\alpha=0.01,\\
32.494 + 3.05543  \log\mid\delta L-\delta L_c\mid, & $for$~Q=0.094984,\alpha=0.02.\\
\end{cases}
\end{equation}
It is obvious that the slope of the fitted straight line is also about 3, which implies that the critical exponent of the
analogous heat capacity  is -2/3 in the $T-\delta L$ plane. The phase structures shown by equal time two heavy operators correlation function is consistent with that given by the holographic entanglement entropy and expectation value of Wilson loop.

\section{Conclusions}
In this paper, we have studied the thermal entropy of a (4+1)-dimensional spherical Gauss-Bonnet-AdS  black hole and found that there
is van der Waals-like phase transition in the $T-S$ plane for a fixed charge $Q$  or a fixed Gauss-Bonnet parameter $\alpha$. For the case $Q=0$, the neural spherical Gauss-Bonnet-AdS  black hole still undergos the van der Waals-like phase transition rather than the Hawking-Page phase transition appeared in the Einstein gravity, which have been studied intensively.

For spherical AdS black hole in Einstein gravity, \cite{Johnson} has observed that holographic entanglement entropy will also undergo van der Waals-like phase transition which is much analogous to that in thermal entropy. We  extended this observation to (4+1)-dimensional spherical Gauss-Bonnet-AdS  black hole and found there is a similar phenomenon. We apply AdS/CFT to study some non-local observables such as holographic entanglement entropy, Wilson loop, and  two point correlation function, which are dual to the  minimal volume, minimal area, and geodesic length respectively in the conformal field theory. For a fixed charge or a fixed Gauss-Bonnet parameter, all these quantities show that there exist van der Waals-like phase transitions, which happen in thermal entropy already.  Below the critical charge or critical Gauss-Bonnet parameter, there exists phase which is composed by a small black hole, large black hole and a intermediate black hole. In this phase, the intermediate black hole will be not stable and the small black hole will directly jump to the large black hole as the temperature increases to the first order phase transition temperature $T_f$.  More precisely, we checked Maxwell's equal area law and found it was valid for all the charges and Gauss-Bonnet parameters to confirm the first order phase transition. As the value of the charge or Gauss-Bonnet parameter increases to the critical value, the small black hole and the large black hole merges into one and the intermediate black hole disappear at transition temperature $T_c$. For this case, phase transition will be the second order. The critical exponent of the analogous heat capacity  is found to be consistent with that of the mean field theory. The black hole is always stable as the value of the charge or Gauss-Bonnet parameter is larger than the critical value. Our results confirm the fact  that all the nonlocal quantities  exhibit van der Waals-like phase transitions in the dual field theory regardless of the dual gravity model.

\section*{Acknowledgements}

We would like to thank Rong-Gen Cai for his discussions.  S.H. is supported by Max-Planck fellowship in Germany and the National Natural Science Foundation of China (No.11305235). L.L. is supported  by the National Natural Science Foundation of China (Grant Nos. 11575270).
X.Z. is supported  by the National Natural Science Foundation of China (Grant Nos. 11405016), China Postdoctoral Science Foundation (Grant No. 2016M590138), Natural Science
Foundation of  Education Committee of Chongqing (Grant No. KJ1500530), and Basic Research Project of Science and Technology Committee of Chongqing(Grant No. cstc2016jcyja0364).

\end{document}